\documentclass[bibyear]{aa} 
\usepackage{color}
\usepackage{graphicx}
\usepackage{txfonts}
\usepackage{natbib}
\bibpunct{}{}{,}{a}{}{,} 
\usepackage{color}

\begin{document}

%
\def\hi {H\,{\sc i}}
\def\hii {H\,{\sc ii}}
\def\water {H$_2$O}
\def\meth {CH$_{3}$OH}
\def\dg{$^{\circ}$}
\def\kms{km\,s$^{-1}$}
\def\ms{m\,s$^{-1}$}
\def\jyb{Jy\,beam$^{-1}$}
\def\mjyb{mJy\,beam$^{-1}$}
\def\solmass {\hbox{M$_{\odot}$}}
\def\solum {\hbox{L$_{\odot}$}} 
\def\d {$^{\circ}$}
\def\n {$n_{\rm{H_{2}}}$}
\def\kmsg{km\,s$^{-1}$\,G$^{-1}$}
\def\tbo {$T_{\rm{b}}\Delta\Omega$}
\def\tb {$T_{\rm{b}}$}
\def\om{$\Delta\Omega$}
\def\dvi {$\Delta V_{\rm{i}}$}
\def\dvz {$\Delta V_{\rm{Z}}$}
\def\code {FRTM code}
\title{The magnetic field at milliarcsecond resolution around \object{IRAS20126+4104}.}

\author{G.\ Surcis  \inst{1}
  \and 
  W.H.T. \ Vlemmings \inst{2}
 \and
  H.J. \ van Langevelde \inst{1,3}
 \and
  L. \ Moscadelli \inst{4}
  \and
  B. \ Hutawarakorn Kramer \inst{5,6}
  }

\institute{
 Joint Institute for VLBI in Europe, Postbus 2, 79990 AA Dwingeloo, The Netherlands
 \and
 Chalmers University of Technology, Onsala Space Observatory, SE-439 92 Onsala, Sweden
 \and
 Sterrewacht Leiden, Leiden University, Postbus 9513, 2300 RA Leiden, The Netherlands
 \and
  INAF-Osservatorio Astrofisico di Arcetri, Largo E. Fermi 5, 50125 Firenze, Italy
 \and
 Max-Planck Institut f\"{u}r Radioastronomie, Auf dem H\"{u}gel 69, 53121 Bonn, Germany
 \and
 National Astronomical Research Institute of Thailand, Ministry of Science and Technology, Rama VI Rd., Bangkok 10400, Thailand
  }

\date{Received ; accepted}
\abstract
{IRAS20126+4104 is a well studied B0.5 protostar that is surrounded by a $\sim$1000~au Keplerian disk and is where a large-scale
outflow originates. Both 6.7-GHz \meth ~masers and 22-GHz \water ~masers have been detected toward this young stellar object. 
The \meth ~masers trace the Keplerian disk, while the \water ~masers are associated with the surface of the conical jet.
Recently, observations of dust polarized emission (350~$\mu$m) at an angular resolution of 9 arcseconds ($\sim15000$~au) have 
revealed an $S$-shaped morphology of the magnetic field around IRAS20126+4104.}
{The observations of polarized maser emissions at milliarcsecond resolution ($\sim20$~au) can make a crucial contribution 
to understanding the orientation of the magnetic field close to IRAS20126+4104. This will allow us to determine whether the magnetic field
morphology changes from arcsecond resolution to milliarcsecond resolution.}
{The European VLBI Network was used to measure the linear polarization and the Zeeman splitting of the 6.7-GHz \meth ~masers toward 
IRAS20126+4104. The NRAO Very Long Baseline Array was used to measure the linear polarization and the Zeeman splitting of the 22-GHz \water 
~masers toward the same region.}
{We detected 26 \meth ~masers and 5 \water ~masers at high angular resolution. Linear polarization emission was observed toward 
three \meth ~masers and toward one \water ~maser. Significant Zeeman splitting was measured in one \meth ~maser ($\Delta V_{\rm{Z}}=-9.2\pm1.4$~\ms).
 No significant (5$\sigma$) magnetic field strength was measured using the \water ~masers. We found that in IRAS20126+4104 the 
rotational energy is less than the magnetic energy.  }
{}


\maketitle
\section{Introduction}
In the past years, the formation of high-mass stars has been at the center of numerous studies, both observational and theoretical.
The observations reveal that the structure of massive protostars is probably similar to that of their less massive counterpart (e.g.,
Tang et al. \cite{tan09}; Keto \& Zhang \cite{ket10}; Johnston et al. \cite{joh13}), and the theoretical simulations match the 
observations as long as the magnetic field is taken into consideration (e.g., Peters et al. \cite{pet11}; Seifried et al. \cite{sei12a};
 Myers et al. \cite{mye13}).\\
\indent One of the typical characteristics of low-mass protostars that has also been observed around high-mass protostars (B-type stars) 
is the presence of circumstellar disks (e.g., Cesaroni et al. \cite{ces06, ces07}). Seifried et al. (\cite{sei11}) show that Keplerian 
disks with sizes of a few 100~au are easily formed  around massive protostars when a weak magnetic field is considered in the simulations. 
The Keplerian disks are also formed if a strong magnetic field is present but only if a turbulent velocity field is introduced
 (Seifried et al. \cite{sei12b}). \\
\indent Determining the morphology of magnetic fields close to circumstellar disks or tori in the early stages of massive star formation 
is very difficult mainly because the massive protostars are distant, rare, and quick to evolve. However, it was possible in some cases, for instance in 
Cepheus~A (Vlemmings et al. \cite{vle10}) and in NGC7538 (Surcis et al. \cite{sur11a}), where the 6.7-GHz \meth ~maser emission was used to
probe the magnetic field at milliarcsecond (mas) resolution (i.e., $\sim10$~au). In both cases, the masers trace the infalling gas but not the disk/torus material 
directly. A suitable case where the magnetic field can be measured on the surface of a disk may instead be IRAS20126+4104.\\
\indent IRAS20126+4104 is a well studied B0.5 protostar ($M=7$~\solmass) at a distance of $1.64\pm0.05$~kpc 
(Moscadelli et al. \cite{mos11}, hereafter MCR11). A disk of $\sim1000$~au ($\rm{PA_{disk}}=53^{\circ}\pm7^{\circ}$, Cesaroni et al. \cite{ces05}), 
which is undergoing Keplerian rotation, was imaged by Cesaroni et al. (\cite{ces97, ces99, ces05}). In addition, a jet/outflow perpendicular 
to the disk ($\rm{PA_{jet}}=115^{\circ}$, MCR11), which shows a precession motion around the rotation axis of the disk (e.g., Shepherd 
et al. \cite{she00}),  was also detected from small- ($\sim10^2$~au) to large-scale ($\sim10^4$~au) (e.g., Cesaroni et al. \cite{ces97, ces99, ces13};
 Hofner et al. \cite{hof07}; Caratti o Garatti \cite{car08}; MCR11). The three maser species 6.7-GHz \meth, 1.6-GHz OH, and 22-GHz \water
~were detected (Edris et a. \cite{edr05}; Moscadelli et al. \cite{mos05}; MCR11). The former can be divided into two groups, i.e. Groups~1 and~2. 
While Group~1 is associated to the Keplerian disk, Group~2 shows relative proper motions, indicating that the masers are moving perpendicularly 
away from the disk (MCR11). The OH masers have an elongated distribution and trace part of the Keplerian disk (Edris et al. \cite{edr05}).
Edris et al. (\cite{edr05}) also identified one Zeeman pair of OH masers that indicates a magnetic field strength of $\sim+11$~mG.
The \water ~masers are instead associated with the surface of the conical jet (opening angle~$=9^{\circ}$), 
with speed increasing for increasing distance from the protostar (Moscadelli et al. \cite{mos05}; MCR11).\\
\indent Shinnaga et al. (\cite{shi12}) measured the polarized dust emission at 350~$\rm{\mu m}$ at arcsecond resolution ($\sim10^4$~au) by using the 
SHARC II Polarimeter (SHARP) with the 10.4~m Leighton telescope at the Caltech Submillimeter Observatory (CSO). They determined that the global 
magnetic field is oriented north-south, but it changes its direction close to the protostar becoming parallel to the Keplerian disk; i.e., here the 
field is nearly perpendicular to the rotation axis of the disk. The apparent jet precession could be explained by the misalignment of the magnetic field and 
the rotation axis (Shinnaga et al. \cite{shi12}). \\
\indent The observations of polarized emissions of 6.7-GHz \meth ~and 22-GHz \water ~masers offer a possibility to better determine the morphology 
of the magnetic field close to the circumstellar disk and to the jet. For this reason, here we present both European VLBI Network (EVN) observations
of \meth ~masers and Very Long Baseline Array (VLBA) observations of \water ~masers that were carried on in full polarization mode.
\begin {table*}[t!]
\caption []{Parameters of the 6.7-GHz \meth ~maser features detected in IRAS20126+4104.} 
\begin{center}
\scriptsize
\begin{tabular}{ l c c c c c c c c c c c c c}
\hline
\hline
\,\,\,\,\,(1)&(2)    & (3)      & (4)      & (5)            & (6)            & (7)                 & (8)         & (9)       & (10)                    & (11)                        & (12)         &(13)                  &(14)           \\
Maser     &Group&RA\tablefootmark{a}&Dec\tablefootmark{a}& Peak flux & $V_{\rm{lsr}}$& $\Delta v\rm{_{L}}$ &$P_{\rm{l}}\tablefootmark{b}$ &  $\chi\tablefootmark{b}$   & $\Delta V_{\rm{i}}\tablefootmark{c}$ & $T_{\rm{b}}\Delta\Omega\tablefootmark{c}$& $P_{\rm{V}}$ & $\Delta V_{\rm{Z}}$  &$\theta\tablefootmark{d}$\\
          &          &offset    & offset   & density(I)     &                &                     &             &            &                         &                             &              &                      &      \\ 
          &          &(mas)     & (mas)    & (Jy/beam)      &  (km/s)        &      (km/s)         & (\%)        &   (\d)    & (km/s)                  & (log K sr)                  &   ($\%$)     &  (m/s)               &(\d)       \\ 
\hline
M01       & 2        & -14.869  & 5.734    & $0.917\pm0.003$&  -6.72         &      $0.29$         & $-$         & $-$       &  $-$                    & $-$                         & $-$                 & $-$                  &$-$ \\ 
M02       & 2        & -11.405  & 3.063    & $0.964\pm0.003$&  -6.72         &      $0.36$         & $-$         & $-$       &  $-$                    & $-$                         & $-$                 & $-$                  &$-$ \\ 
M03       & 2        & -2.797   & 19.127   & $0.265\pm0.009$&  -6.10         &      $0.20$         & $-$         & $-$       &  $-$                    & $-$                         & $-$                 & $-$                  &$-$ \\ 
M04       & 2        & -1.743   & -16.438  & $0.275\pm0.007$&  -5.97         &      $0.25$         & $-$         & $-$       &  $-$                    & $-$                         & $-$                 & $-$                  &$-$ \\ 
M05       & 2        & 0        & 0        & $27.838\pm0.009$& -6.10         &      $0.36$         & $1.6\pm0.4$ & $-65\pm3$ &  $2.0^{+0.1}_{-0.2}$    & $8.8^{+0.8}_{-0.4}$         & $0.6$        & $-9.2\pm1.4$         &$75^{+10}_{-43}$ \\ 
M06       & 2        & 0.129    & -7.450   & $0.316\pm0.007$&  -6.14         &      $0.30$         & $-$         & $-$       &  $-$                    & $-$                         & $-$                 & $-$                  &$-$ \\ 
M07       & 2        & 0.947    & 8.122    & $0.543\pm0.008$&  -6.01         &      $0.31$         & $-$         & $-$       &  $-$                    & $-$                         & $-$                 & $-$                  &$-$ \\ 
M08       & 2        & 9.382    & -4.685   & $0.428\pm0.007$&  -5.97         &      $0.25$         & $-$         & $-$       &  $-$                    & $-$                         & $-$                 & $-$                  &$-$ \\ 
M09       & 2        & 19.883   & -7.031   & $0.199\pm0.007$&  -6.23         &      $0.20$         & $-$         & $-$       &  $-$                    & $-$                         & $-$                 & $-$                  &$-$ \\ 
M10       & 2        & 19.904   & -11.261  & $0.049\pm0.003$&  -5.66         &      $0.23$         & $-$         & $-$       &  $-$                    & $-$                         & $-$                 & $-$                  &$-$ \\ 
M11       & 2        & 52.634   & -48.145  & $0.068\pm0.002$&  -5.13         &      $0.28$         & $-$         & $-$       &  $-$                    & $-$                         & $-$                 & $-$                  &$-$ \\ 
M12       & 2        & 56.313   & -21.915  & $0.308\pm0.002$&  -5.57         &      $0.38$         & $-$         & $-$       &  $-$                    & $-$                         & $-$                 & $-$                  &$-$ \\ 
M13       & 2        & 62.231   & -15.331  & $0.861\pm0.003$&  -6.41         &      $0.27$         & $-$         & $-$       &  $-$                    & $-$                         & $-$                 & $-$                  &$-$ \\ 
M14       & 2        & 81.877   & -10.986  & $0.919\pm0.003$&  -6.67         &      $0.28$         & $-$         & $-$       &  $-$                    & $-$                         & $-$                 & $-$                  &$-$ \\ 
M15       &          & 83.835   & 244.766  & $0.070\pm0.003$&  -6.50         &      $0.19$         & $-$         & $-$       &  $-$                    & $-$                         & $-$                 & $-$                  &$-$ \\ 
M16       & 1        & 166.917  & -77.072  & $0.048\pm0.002$&  -5.18         &      $0.19$         & $-$         & $-$       &  $-$                    & $-$                         & $-$                 & $-$                  &$-$ \\ 
M17       & 1        & 155.448  & -104.588 & $0.178\pm0.002$&  -4.87         &      $0.23$         & $-$         & $-$       &  $-$                    & $-$                         & $-$                 & $-$                  &$-$ \\ 
M18       & 1        & 191.556  & 37.796   & $1.851\pm0.003$&  -7.64         &      $0.27$         & $1.4\pm0.1$ & $-71\pm5$ &  $1.4^{+0.2}_{-0.2}$    & $8.8^{+0.3}_{-0.1}$         & $-$           & $-$                  &$85^{+6}_{-34}$ \\ 
M19       & 1        & 192.782  & 25.593   & $0.117\pm0.003$&  -7.68         &      $0.23$         & $-$         & $-$       &  $-$                    & $-$                         & $-$                 & $-$                  &$-$ \\ 
M20       & 1        & 204.295  & 7.683    & $0.309\pm0.003$&  -7.11         &      $0.21$         & $-$         & $-$       &  $-$                    & $-$                         & $-$                 & $-$                  &$-$ \\ 
M21       & 1        & 207.436  & 33.264   & $0.072\pm0.003$&  -7.51         &      $0.19$         & $-$         & $-$       &  $-$                    & $-$                         & $-$                 & $-$                  &$-$ \\ 
M22       & 1        & 210.965  & 16.399   & $0.114\pm0.003$&  -6.50         &      $0.21$         & $-$         & $-$       &  $-$                    & $-$                         & $-$                 & $-$                  &$-$ \\ 
M23       & 1        & 215.312  & 8.267    & $0.402\pm0.003$&  -6.98         &      $0.36$         & $-$         & $-$       &  $-$                    & $-$                         & $-$                 & $-$                  &$-$ \\ 
M24       & 1        & 237.153  & 4.383    & $2.154\pm0.003$&  -6.98         &      $0.30$         & $0.6\pm0.2$ & $56\pm43$ &  $1.6^{+0.2}_{-0.2}$    & $8.4^{+0.4}_{-0.4}$         & $-$           & $-$                  &$79^{+11}_{-37}$ \\ 
M25       & 1        & 261.232  & 4.707    & $0.166\pm0.003$&  -7.72         &      $0.28$         & $-$         & $-$       &  $-$                    & $-$                         & $-$                 & $-$                  &$-$ \\ 
M26       & 1        & 277.392  & 3.834    & $0.644\pm0.002$&  -8.25         &      $0.26$         & $-$         & $-$       &  $-$                    & $-$                         & $-$                 & $-$                  &$-$ \\ 
\hline
\end{tabular} \end{center}
\tablefoot{
\tablefoottext{a}{The reference position is $\alpha_{2000}=20^{\rm{h}}14^{\rm{m}}26^{\rm{s}}\!.046\pm0^{\rm{s}}\!.001$ and 
$\delta_{2000}=41^{\circ}13'32''\!\!.690\pm0''\!\!.009$ (see Sect.~\ref{res}).}
\tablefoottext{b}{$P_{\rm{l}}$ and $\chi$ are the mean values of the linear polarization fraction and the linear polarization angle measured across the spectrum, respectively.}
\tablefoottext{c}{The best-fitting results obtained by using a model based on the radiative transfer theory of methanol masers 
for $\Gamma+\Gamma_{\nu}=1~\rm{s^{-1}}$ (Vlemmings et al. \cite{vle10}, Surcis et al. \cite{sur11a}). The errors were determined 
by analyzing the full probability distribution function.}
\tablefoottext{d}{The angle between the magnetic field and the maser propagation direction is determined by using the observed $P_{\rm{l}}$ 
and the fitted emerging brightness temperature. The errors were determined by analyzing the full probability distribution function.}
}
\label{IRAS20_meth}
\end{table*}
\section{Observations}
\subsection{6.7-GHz EVN data}
IRAS20126+4104 was observed at 6.7-GHz in full polarization spectral mode with seven of the EVN\footnote{The European VLBI 
Network is a joint facility of European, Chinese, South African, and other radio astronomy institutes funded by their national research 
councils.} antennas (Effelsberg, Jodrell, Onsala, Medicina, Torun, Westerbork, and Yebes-40\,m), for a total observation time of 5.5~h, on 
October 30, 2011 (program code ES066). The bandwidth was 2~MHz, providing a 
velocity range of $\sim100$~\kms. The data were correlated with the EVN software correlator (SFXC) at the Joint Institute for VLBI in Europe
 (JIVE) using 2048 channels and generating all four polarization combinations (RR, LL, RL, LR) with a spectral resolution of $\sim$1~kHz 
($\sim$0.05~\kms).\\
\indent The data were edited and calibrated using the Astronomical Image Processing System (AIPS). The bandpass, delay, phase, and 
polarization calibration were performed on the calibrator J2202+4216. Fringe-fitting and 
self-calibration were performed on the brightest maser feature (M05 in Table \ref{IRAS20_meth}). Then the \textit{I}, \textit{Q}, \textit{U}, and 
\textit{V} cubes were imaged ($\rm{rms}=2.4$~\mjyb) using the AIPS task IMAGR. The beam size was 7.47~mas~$\times$~3.38~mas
($\rm{PA}=76$\d). The \textit{Q} and \textit{U} cubes were combined to produce 
cubes of polarized intensity 
($POLI=\sqrt{Q^{2}+U^{2}}$) and polarization angle ($\chi=1/2\times~atan(U/Q)$). We calibrated the linear polarization angles by comparing 
the linear polarization angle of the polarization calibrator measured by us with the angle obtained by calibrating the POLCAL observations 
made by NRAO\footnote{http://www.aoc.nrao.edu/$\sim$~smyers/calibration/}. IRAS20126+4104 was observed between two POLCAL observations runs during
 which the linear polarization angle of J2202+4216 was constant, with an average value of $-31^{\circ}\pm1$\d. We were therefore able to estimate the 
polarization angle with a systemic error of no more than $\sim$~1\d. The formal errors on $\chi$ are due to thermal noise. This error is given by 
$\sigma_{\chi}=0.5 ~\sigma_{P}/P \times 180^{\circ}/\pi$ (Wardle \& Kronberg \cite{war74}), where $P$ and $\sigma_{P}$ are the polarization 
intensity and corresponding rms error, respectively.
\begin {table*}[t!]
\caption []{Parameters of the 22-GHz \water ~maser features detected in IRAS20126+4104.} 
\begin{center}
\scriptsize
\begin{tabular}{ l c c c c c c c c c c c c}
\hline
\hline
\,\,\,\,\,(1)&(2)    & (3)      & (4)            & (5)            & (6)                 & (7)         & (8)       & (9)                     & (10)                        & (11)         & (12)                 &(13)                       \\
Maser     & RA\tablefootmark{a}& Dec\tablefootmark{a}& Peak flux   & $V_{\rm{lsr}}$ & $\Delta v\rm{_{L}}$ &$P_{\rm{l}}\tablefootmark{b}$ &  $\chi\tablefootmark{b}$   & $\Delta V_{\rm{i}}$\tablefootmark{c} & $T_{\rm{b}}\Delta\Omega$\tablefootmark{c}& $P_{\rm{V}}$ & $\Delta V_{\rm{Z}}$  & $\theta^{d}$\\
          & offset   & offset   & density(I)     &                &                     &             &            &                         &                             &              &                      &       \\ 
          & (mas)    & (mas)    & (Jy/beam)      &  (km/s)        &      (km/s)         & (\%)        &   (\d)    & (km/s)                  & (log K sr)                  &   ($\%$)     &  (m/s)               &  (\d)       \\ 
\hline
W01       & -0.818   & -0.656   & $0.37\pm0.05$  & -2.05          &      $0.74$         & $-$         & $-$       &  $-$                    & $-$                         & $-$            & $-$                  & $-$    \\ 
W02       & 0        &  0       & $24.77\pm0.06$ & -4.61          &      $1.37$         & $1.3\pm0.2$ & $-37\pm13$&  $<0.5$                 & $9.1^{+0.3}_{-0.4}$         & $-$            & $-$                  & $90^{+9}_{-9}$    \\ 
W03       & 403.317  & -212.020 & $0.19\pm0.04$  & -5.61          &      $0.73$         & $-$         & $-$       &  $-$                    & $-$                         & $-$            & $-$                  & $-$    \\ 
W04       & 403.898  & -212.452 & $0.39\pm0.05$  & -6.23          &      $0.55$         & $-$         & $-$       &  $-$                    & $-$                         & $-$            & $-$                  & $-$     \\ 
W05       & 542.648  & -201.458 & $0.23\pm0.05$  & -15.51         &      $0.39$         & $-$         & $-$       &  $-$                    & $-$                         & $-$            & $-$                  & $-$     \\ 
\hline
\end{tabular} \end{center}
\tablefoot{
\tablefoottext{a}{The reference position is $\alpha_{2000}=20^{\rm{h}}14^{\rm{m}}25^{\rm{s}}\!.966\pm0^{\rm{s}}\!.002$ and 
$\delta_{2000}=41^{\circ}13'32''\!\!.738\pm0''\!\!.014$ (see Sect.~\ref{res}).}
\tablefoottext{b}{$P_{\rm{l}}$ and $\chi$ are the mean values of the linear polarization fraction and the linear polarization angle measured across the spectrum, respectively.}
\tablefoottext{c}{The best-fitting results obtained by using a model based on the radiative transfer theory of \water ~masers 
for $\Gamma+\Gamma_{\nu}=1~\rm{s^{-1}}$ (Surcis et al. \cite{sur11b}). The errors were determined 
by analyzing the full probability distribution function.}
\tablefoottext{d}{The angle between the magnetic field and the maser propagation direction is determined by using the observed $P_{\rm{l}}$ 
and the fitted emerging brightness temperature. The errors were determined by analyzing the full probability distribution function.}
}
\label{IRAS20_wat}
\end{table*}
\begin{figure*}[th!]
\centering
\includegraphics[width = 8 cm]{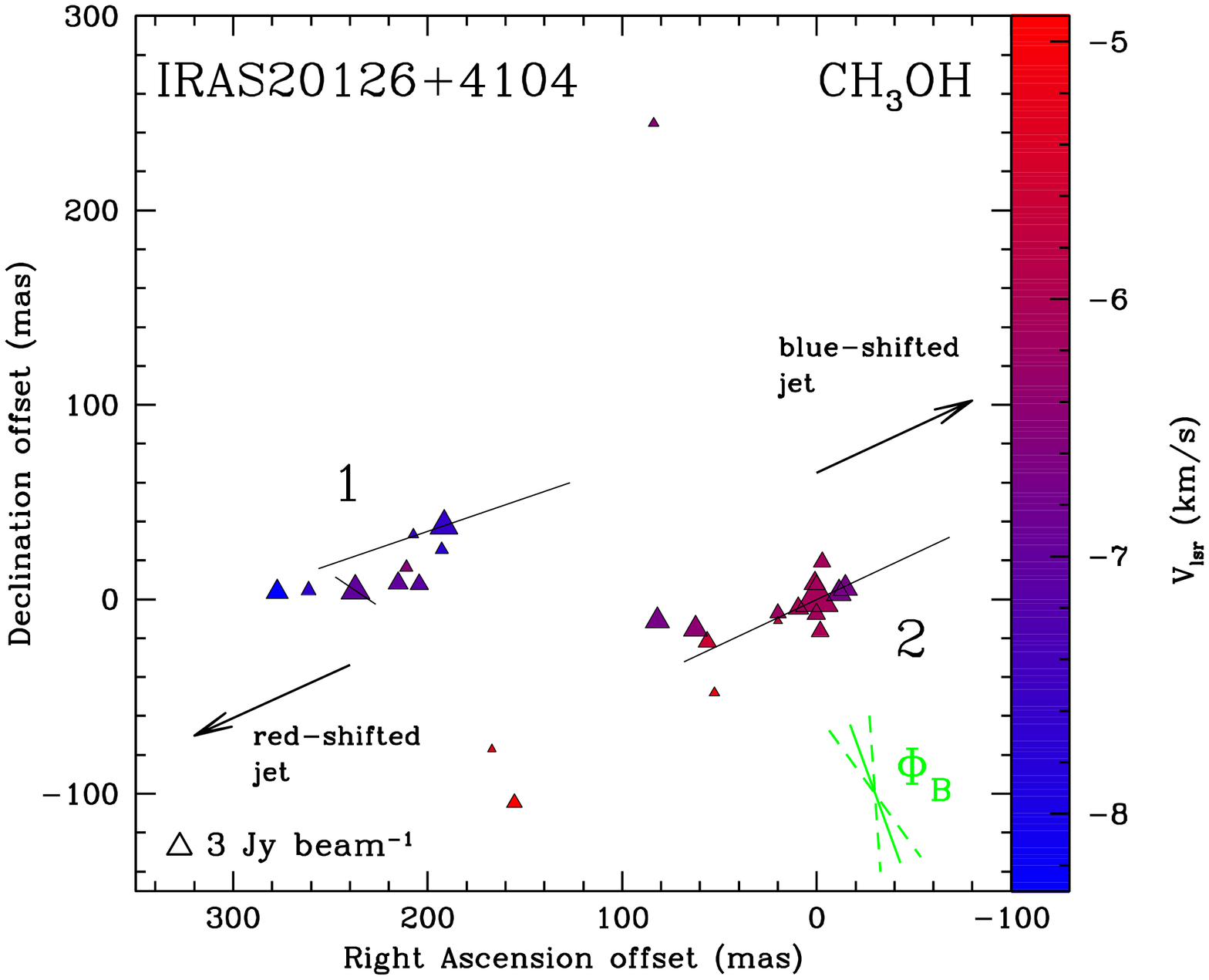}
\includegraphics[width = 8 cm]{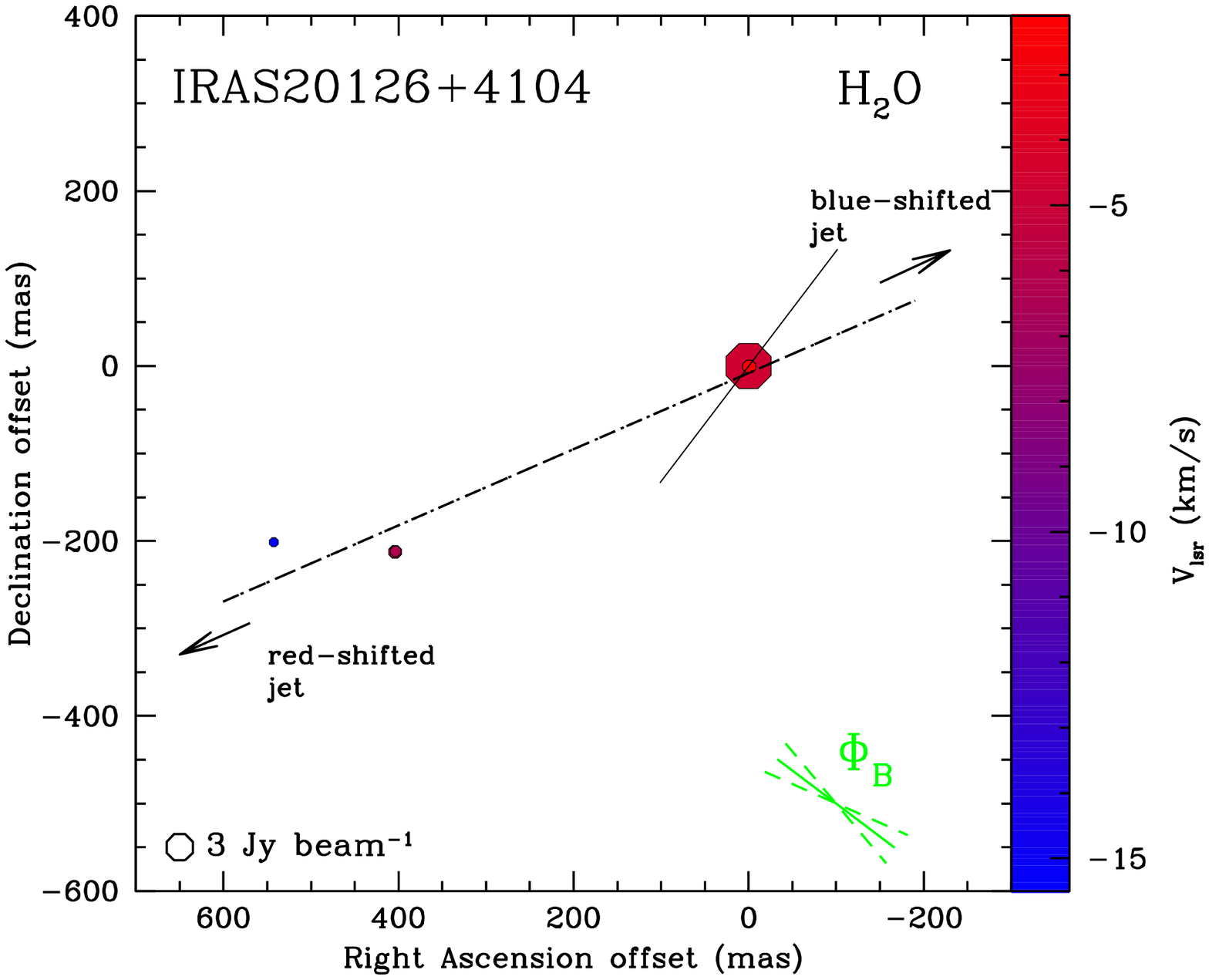}
\caption{Left panel: a view of the 6.7-GHz \meth ~maser features detected around IRAS20126+4104, the reference position is 
$\alpha_{2000}=20^{\rm{h}}14^{\rm{m}}26^{\rm{s}}\!.046$ and $\delta_{2000}=41^{\circ}13'32''\!\!.690$ (see Sect.~\ref{res}). 
Right panel: a view of the 22-GHz \water ~maser features detected around IRAS20126+4104, the reference position is 
$\alpha_{2000}=20^{\rm{h}}14^{\rm{m}}25^{\rm{s}}\!.966$ and $\delta_{2000}=41^{\circ}13'32''\!\!.738$ (see Sect.~\ref{res}).
The triangles and the octagonal symbols are the identified \meth ~and \water ~maser features, respectively, scaled logarithmically according to their peak
flux density (Tables~\ref{IRAS20_meth} and \ref{IRAS20_wat}). The maser LSR radial velocity is indicated by color. (The systemic velocity of IRAS20126+4104
 is $V_{\rm{lsr}}=-3.5$~\kms, MCR11.) A 3~\jyb ~symbol is plotted for illustration in both
panels. The linear polarization vectors, scaled logarithmically according to polarization fraction $P_{\rm{l}}$, are overplotted. In the right bottom 
corner of both panels, 
the error-weighted orientation of the magnetic field ($\Phi_{\rm{B}}$, see Sect. \ref{magori}) is also reported, the two dashed segments indicate the
 uncertainties. The two arrows indicate the direction but not the absolute position of the red- and blue-shifted lobes of the jet ($\rm{PA_{jet}}=115$\d; 
MCR11). The dotted line is the best linear fit of the \water ~maser features 
($\rm{PA_{H_{2}O}}=114^{\circ}\pm4$\d).}
\label{I20_cp}
\end{figure*}
\subsection{22-GHz VLBA data}
The star-forming region was also observed in the $6_{16}-5_{23}$ transition of \water ~(rest frequency:22.23508~GHz) with the NRAO\footnote{The National 
Radio Astronomy Observatory (NRAO) is a facility of the National Science Foundation operated under cooperative agreement by Associated Universities, Inc.}
 VLBA on June 24, 2012. The observations were made in full polarization mode using a bandwidth of 4~MHz to cover a velocity range of $\sim54$~\kms. The data
were correlated with the DiFX correlator using 2000 channels and generating all four polarization combinations (RR, LL, RL, LR) with a spectral resolution of
2~kHz ($\sim$0.03~\kms). Including the overheads, the total observation time was 8~hr.\\
\indent The data were edited and calibrated using AIPS following the method of Kemball 
et al. (\cite{kem95}). The bandpass, the delay, the phase, and the polarization calibration were performed on the calibrator 
J2202+4216. The fringe-fitting and the self-calibration were performed on the brightest maser feature (W02 in Table \ref{IRAS20_wat}). Then we imaged the
\textit{I}, \textit{Q}, \textit{U}, and \textit{V} cubes ($\rm{rms}=20$~\mjyb) using the AIPS task IMAGR (beam size 
0.75~mas~$\times$~0.34~mas,  $\rm{PA}=-9.4$\d).The \textit{Q} and \textit{U} cubes were combined to produce cubes of \textit{POLI} and $\chi$. Because 
IRAS20126+4104 was observed ten days before a POLCAL observations run, we calibrated the linear polarization angles  of the \water ~masers
by comparing the linear polarization angle of J2202+4216 measured by us with the angles measured during that POLCAL 
observations run ($\chi_{\rm{J2202+4216}}=-15$\d$\!\!.0\pm0$\d$\!\!.3$). Also in the case of the \water ~masers, the $\sigma_{\chi}$ is due to thermal noise.
\begin{figure*}[th!]
\centering
\includegraphics[width = 6.4 cm]{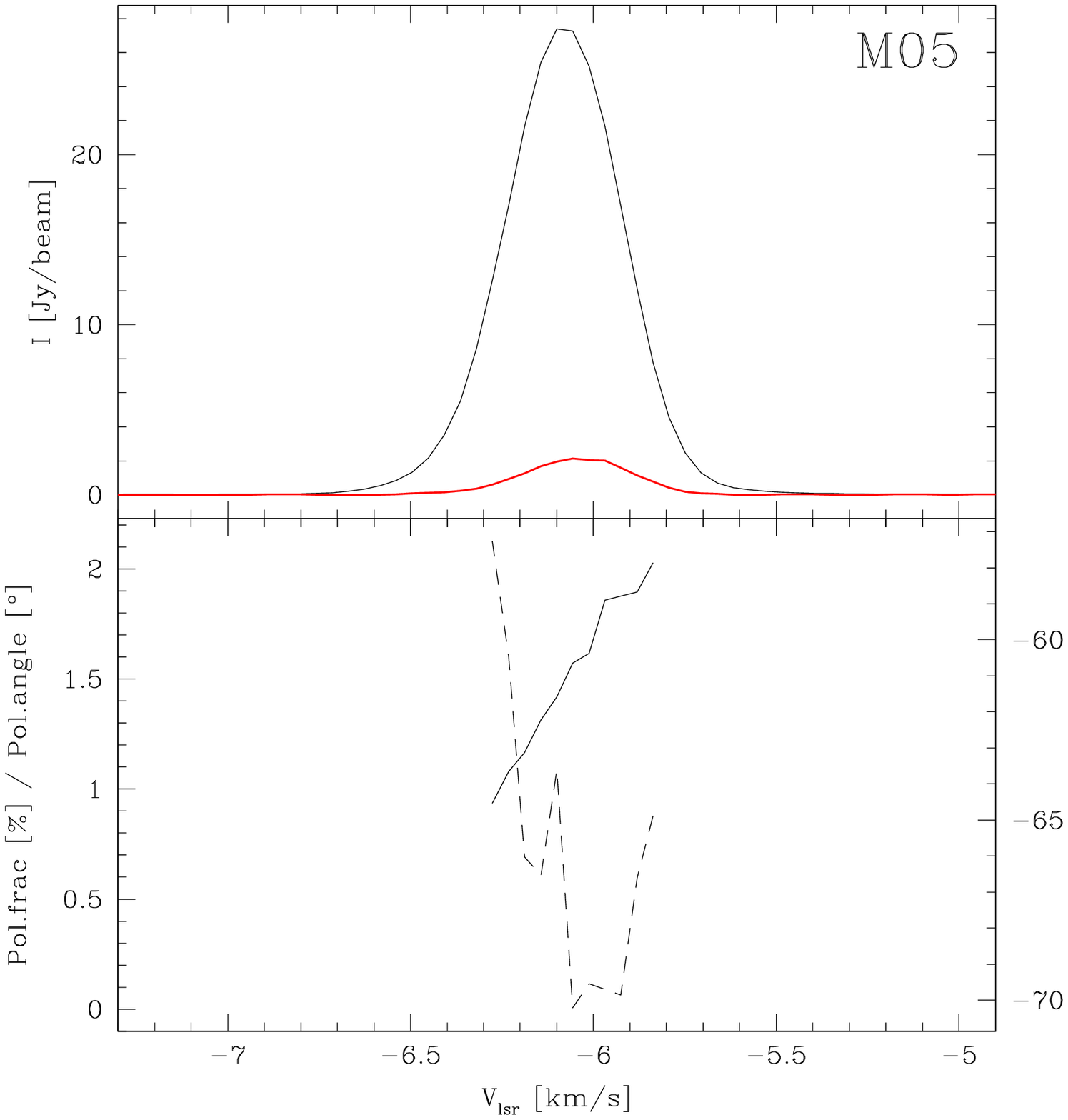}
\includegraphics[width = 6.4 cm]{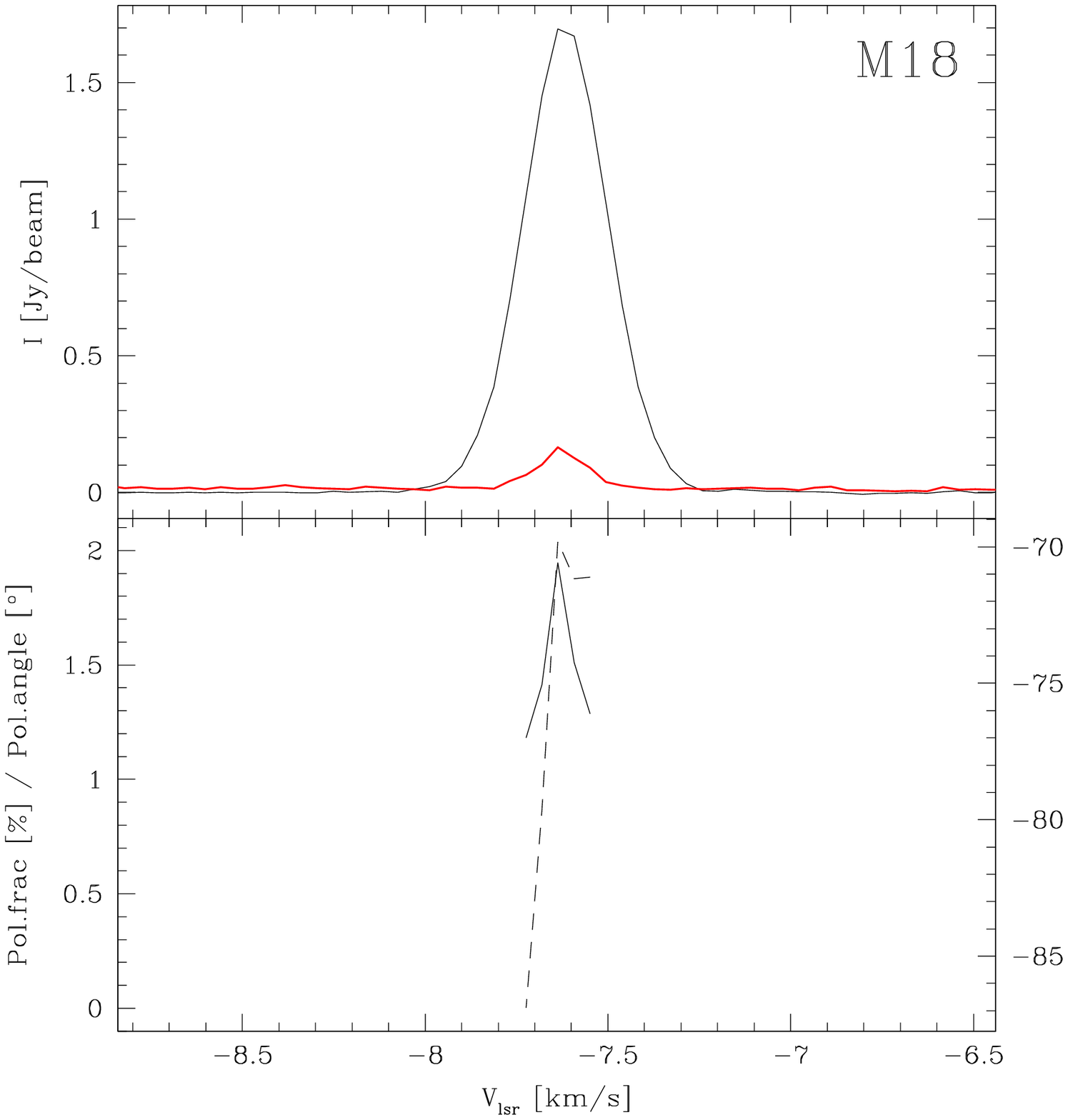}
\includegraphics[width = 6.4 cm]{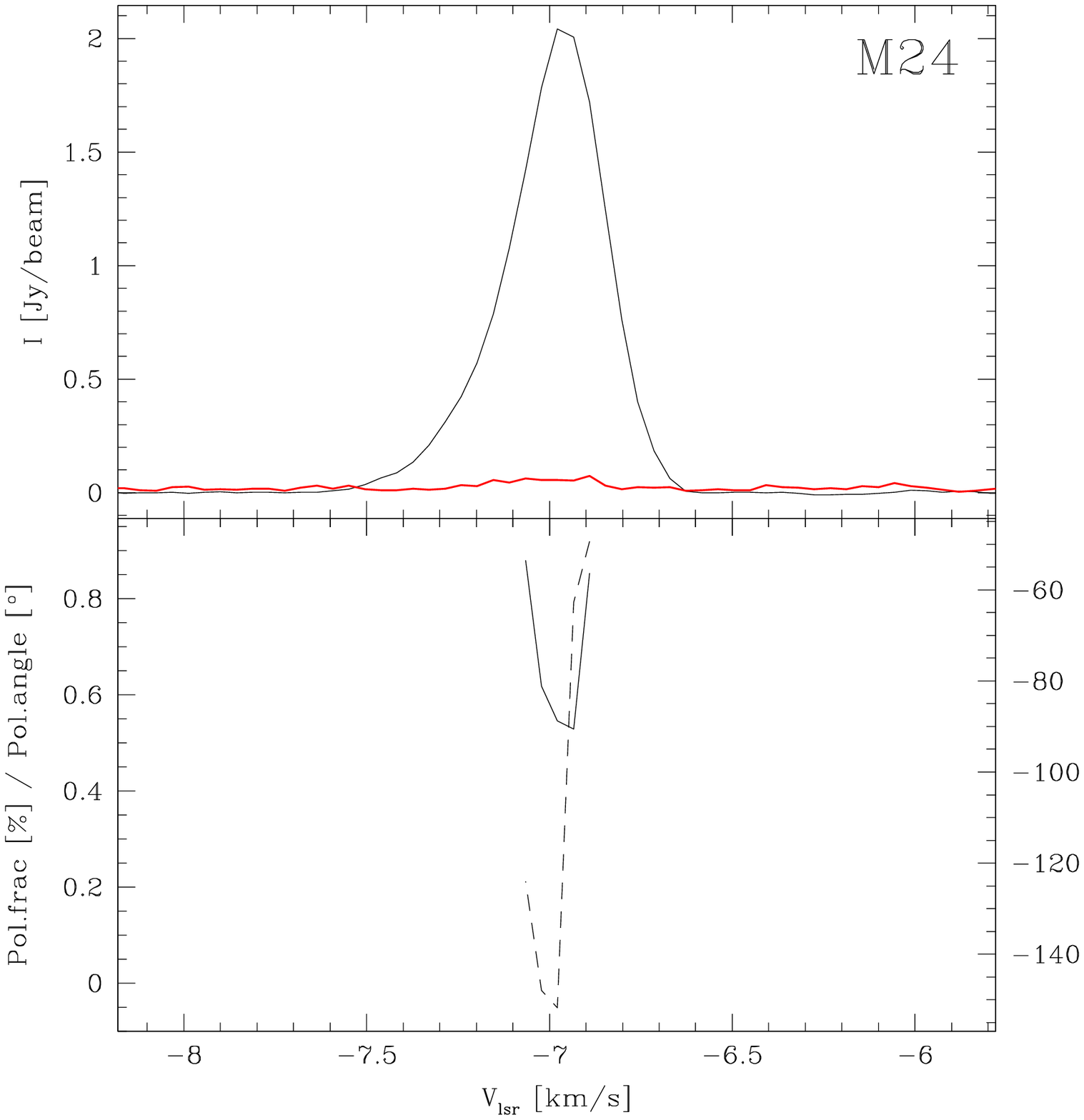}
\includegraphics[width = 6.4 cm]{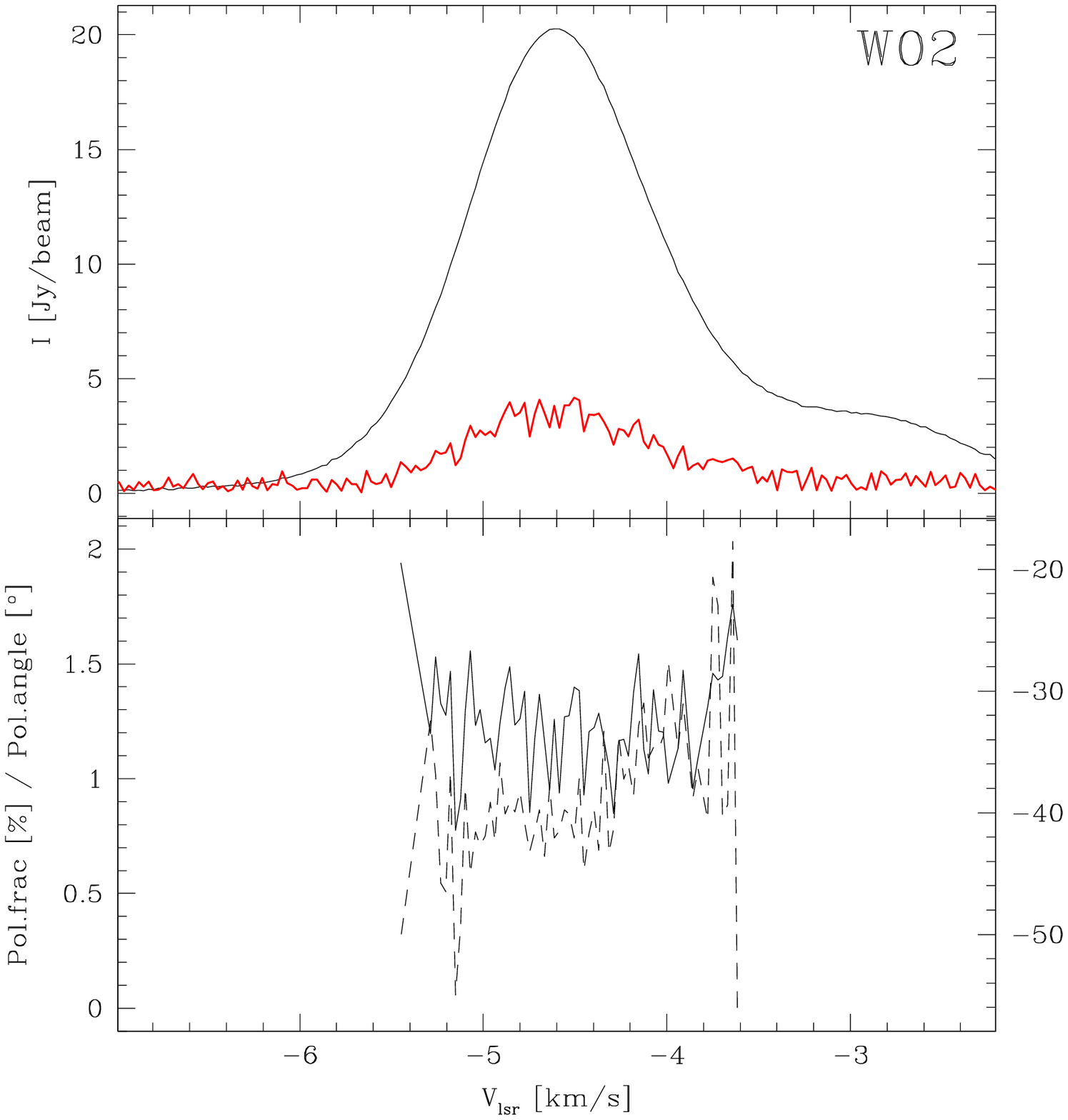}
\caption{Total intensity (\textit{I}, black solid line) and linear polarization intensity (red solid line) spectra of the \meth ~maser features M05, M18, and M24, 
and of the \water ~maser feature W02 (\textit{upper panel}). The linear polarization intensity spectra have been multiplied by a factor of five for the 
M05, M18, and M24, and by a factor of fifteen for W02. The spectra of polarization fraction (black solid line, left scale) and polarization angle (dashed black 
line, right scale) are also shown (\textit{lower panel}).}
\label{POL_plots}
\end{figure*}
\section{Analysis}
\label{analysis}
The \meth ~and \water ~maser features were identified by using the process described in Surcis et al. (\cite{sur11b}). 
We determined the mean linear polarization fraction ($P_{\rm{l}}$) and the mean linear polarization angle ($\chi$) 
of each \meth ~and \water ~maser feature by only considering the consecutive channels (more than two) across the total intensity spectrum for which $POLI\geq5\sigma$.\\
\indent We fitted the total intensity and the 
linearly polarized spectra of \water ~and \meth ~maser features, for which we were able to detect linearly polarized emission, by using the full radiative 
transfer method (FRTM) code for 22-GHz \water ~masers (Vlemmings et al. \cite{vle06}; Surcis et al. \cite{sur11b}) and the adapted version of the code for 
6.7-GHz \meth ~masers (Vlemmings et al. \cite{vle10}; Surcis et al. \cite{sur11a}). The code is based on the models of Nedoluha \&
Watson (\cite{ned92}), who solved the transfer equations for the polarized radiation of 22-GHz \water ~masers in the presence of a magnetic field causing
a Zeeman splitting (\dvz) that is much smaller than the spectral line breadth.\\
\indent We modeled the observed spectra by gridding the intrinsic thermal linewidth (\dvi) in the case of \water ~masers from 0.5 to 3.5~\kms \ in steps
 of 0.025~\kms,  and in the case of the \meth ~masers from 0.5 to 2.4~\kms in steps of 0.05~\kms, by using a least-square fitting routine. The output 
of the codes provides estimates of the emerging brightness temperature (\tbo) and of \dvi. From the fit results, we were able to determine the best estimates 
of the angle between
the maser propagation direction and the magnetic field ($\theta$), because both shape and strength of the linear polarization spectrum depend (nonlinearly)
on the maser saturation level and $\theta$. If $\theta>\theta_{\rm{crit}}=55$\d, where $\theta_{\rm{crit}}$ is the Van Vleck angle, 
the magnetic field appears to be perpendicular to the linear polarization vectors; otherwise, it is parallel (Goldreich et al. \cite{gol73}). To better
determine the orientation of the magnetic field with respect to the linear polarization vectors, Surcis et al. (\cite{sur13}) introduced a method that takes the errors associated to $\theta$  into 
consideration (i.e., $\theta^{\varepsilon^{\rm{+}}}_{\varepsilon^{\rm{-}}}$ in Tables \ref{IRAS20_wat} and \ref{IRAS20_meth}).
We state that if $|\theta^{\rm{+}}-55$\d$|>|\theta^{\rm{-}}-55$\d$|$, where $\theta^{\rm{\pm}}=\theta\pm\varepsilon^{\rm{\pm}}$, 
the magnetic field is most likely  perpendicular to the linear polarization vectors; otherwise, the magnetic field is assumed to be parallel.
 Of course, if $\theta^{\rm{-}}$ and $\theta^{\rm{+}}$ are both larger or smaller than 55\d ~the magnetic field is perpendicular or parallel to the linear 
polarization vectors, respectively.\\
\indent Moreover, the best estimates for \tbo ~and \dvi ~are included in the corresponding code to produce the I and V models that were used for fitting the
 total intensity and circular polarized spectra of the corresponding maser feature.
\section{Results}
\label{res}
Tables~\ref{IRAS20_meth} and 2 list the 26 6.7-GHz \meth ~maser features (named M01--M26) and the 5 22-GHz \water ~maser features
(named W01--W05), respectively, that we detected towards IRAS20126+4104. They are all shown in Fig.~\ref{I20_cp}. Because we did not observe in phase-referencing 
mode, we do not have information for the absolute position of both maser species. Still, we were able to estimate the absolute position of the brightest features
of both maser species (M05 and W02) through fringe rate mapping using the AIPS task FRMAP. The absolute position errors are $\Delta\alpha^{\rm{M05}}=6$~mas and
$\Delta\delta^{\rm{M05}}=9$~mas for the \meth ~maser feature, and $\Delta\alpha^{\rm{W02}}=24$~mas and $\Delta\delta^{\rm{W02}}=14$~mas for the \water ~maser
feature. The position of the brightest \meth ~maser feature M05, which is Feature 1 in MCR11, agrees within 2$\sigma$ with the 
position of 
 Feature 1 after considering the change in position due to the proper motion of the \meth ~masers ($-4~\rm{mas~yr^{-1}}$ both in RA and in DEC, MCR11).\\
\indent The description of the maser distribution and the polarization results are reported for each maser species separately below.
\subsection{\meth ~masers}
The \meth ~maser features can be divided into two groups, 1 and 2, following the naming convention of MCR11. An additional maser feature M15, 
which is undetected by MCR11, is about 200~mas north from the other maser features and cannot be included in any of these two groups.
The spatial distribution and the velocity ranges of the two groups are consistent with those of MCR11.\\
\indent We detected linear polarization in three \meth ~maser features ( $P_{\rm{l}}=0.6\%-1.6\%$, see Fig.~\ref{POL_plots}), and the error-weighted linear
polarization angles is
$\langle\chi\rangle_{\rm{CH_3OH}}=-70^{\circ}\pm16^{\circ}$. The adapted version of the \code ~was able to properly fit 
all these three \meth ~maser features, and the outputs with their relative errors are reported in Cols. 10, 11, and 14 of Table~\ref{IRAS20_meth}. 
Moreover, these maser features appear to be unsaturated, because their \tbo ~are under the saturation threshold 
$(T_{\rm{b}}\Delta\Omega)_{\rm{CH_3OH}}=2.6\times10^9$~K~sr of the 6.7-GHz 
\meth ~masers (Surcis et al. \cite{sur11a}). Considering the determined $\theta$ angles, the magnetic field is perpendicular to the linear polarization 
vectors, i.e., $|\theta^{\rm{+}}-55$\d$|>|\theta^{\rm{-}}-55$\d$|$.
Furthermore, we detected circularly polarized emission ($P_{\rm{V}}=0.6\%$) toward the brightest \meth ~maser feature M05, for which we measured quite a large
Zeeman splitting $\Delta V_{\rm{Z}}=(-9.2\pm1.4)$~\ms.  
\begin{figure}[th!]
\centering
\includegraphics[width = 6.4 cm]{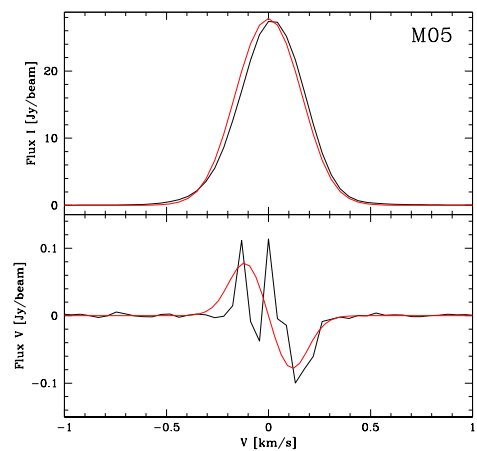}
\caption{Total intensity (\textit{I}, \textit{upper panel}) and circular polarization intensity (\textit{V}, \textit{lower panel}) spectra for the \meth ~maser feature M05.
The thick red line shows the best-fit models of \textit{I} and \textit{V} emission obtained using the adapted FRTM code (see 
Sect.~\ref{analysis}). The maser features were centered on zero velocity.}
\label{Vfit}
\end{figure}
\subsection{\water ~masers}
The \water ~maser features are linearly distributed (PA$_{\rm{H_2O}}=114^{\circ}\pm4$\d) from northwest (NW) to southeast (SE), and their velocities 
 increase in magnitude from NW to SE. The velocity of W05, which is the most southeastern and the most blue-shifted \water ~maser features, is an order 
of magnitude 
faster than the velocities of the other maser features. Although the PA$_{\rm{H_2O}}$ of the maser distribution agrees perfectly with the PA measured 
recently by MCR11, the maser features are not on the outflow as detected by MCR11 and the velocity distribution is reversed with respect to 
what MCR11 observed (see Fig.~\ref{MF}). \\
\indent We detected linearly polarized emission ($P_{\rm{l}}=1.3\%$, see Fig.~\ref{POL_plots}) only from the brightest \water ~maser feature
W02 ($\chi=-37^{\circ}\pm13^{\circ}$). 
The \code ~provides an upper limit of \dvi ~(Col. 9 of Table~\ref{IRAS20_wat}), while the value of \tbo ~(column 10) is below 
the saturation threshold $(T_{\rm{b}}\Delta\Omega)_{\rm{H_2O}}=6.7\times10^9$~K~sr also for the \water ~maser, indicating an unsaturated maser (Surcis et al. \cite{sur11a}). 
The third output of the \code, i.e. $\theta$ (Col. 13), indicates that the magnetic field is on the plane of the sky and perpendicular to the linear 
polarization vector. No circular polarization at $5\sigma$ was detected toward any \water ~maser feature ($P_{\rm{V}}^{\rm{W02}}<0.4\%$).
\section{Discussion}
\subsection{Zeeman splitting}
The magnetic field strength along the line of sight can be calculated from the Zeeman-splitting measurements by using
\begin{equation}
B_{||}=\frac{\Delta V_{\rm{Z}}}{\alpha_{\rm{Z}}},
\end{equation}
where $\alpha_{\rm{Z}}$ is the Zeeman-splitting coefficient, which depends on the Land\'{e} g-factor of the corresponding maser transition. Moreover, the 
total magnetic field strength can be determined if the angle between the maser propagation direction and the magnetic field $\theta$ is known, i.e.,
 $B=B_{||}/cos~\theta$. While the Zeeman-splitting coefficient for the 22-GHz \water ~maser is well-known, $\alpha_{\rm{Z}}$ for the 6.7-GHz \meth ~maser 
emission is still uncertain. Indeed, the Land\'{e} g-factor corresponding to the \meth ~maser transition is still unknown (Vlemmings et al. \cite{vle11}). 
However, a considerable value of $\alpha_{\rm{Z}}^{\rm{CH_3OH}}$ could be in the range $0.005$~\kmsg$<\alpha_{\rm{Z}}^{\rm{CH_3OH}}0.05$~\kmsg ~(Surcis 
et al. \cite{sur11a}).\\
\indent From our observations we measured Zeeman splitting only from the \meth ~maser M05, and consequently we can speculatively give only a possible 
range of $B_{||}$, which is $0.2~\rm{G}<|B^{\rm{CH_3OH}}_{||}|<2.1~\rm{G}$ where the uncertainty of \dvz ~has been taken into account. Considering 
$\theta_{\rm{M05}}=75$\d$^{+10^{\circ}}_{-43^{\circ}}$, the total magnetic field, $B_{\rm{CH_3OH}}$,
 ranges from $-0.2$~G to $-24$~G. According to the sign of the Zeeman splitting, the magnetic field is pointing toward the observer. 
 The non-detection of significant circular polarized emission from the 22-GHz \water ~maser could be due to a weaker magnetic field along the outflows.
\begin{figure*}[t!]
\centering
\includegraphics[width = 8 cm]{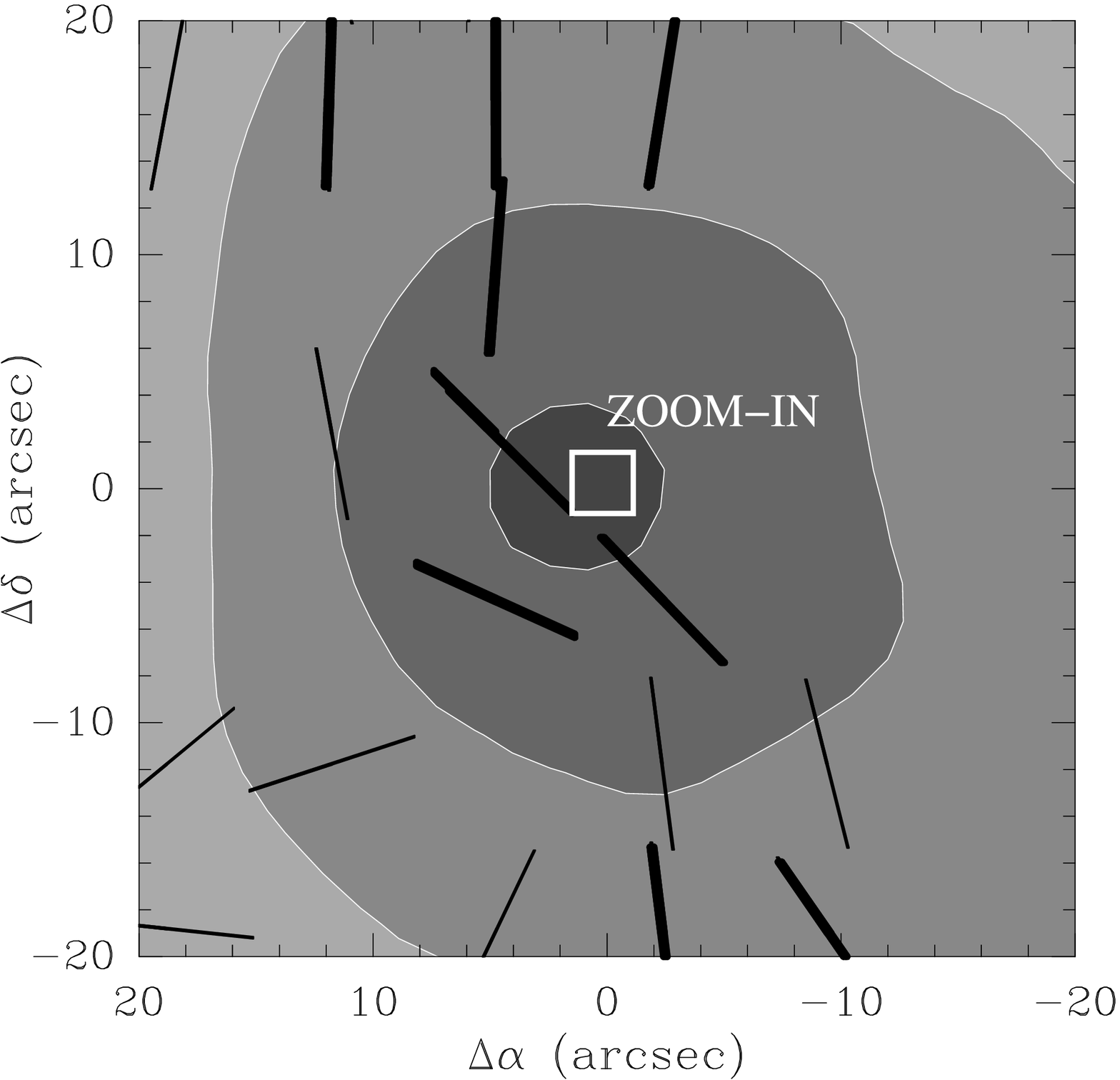}
\includegraphics[width = 8 cm]{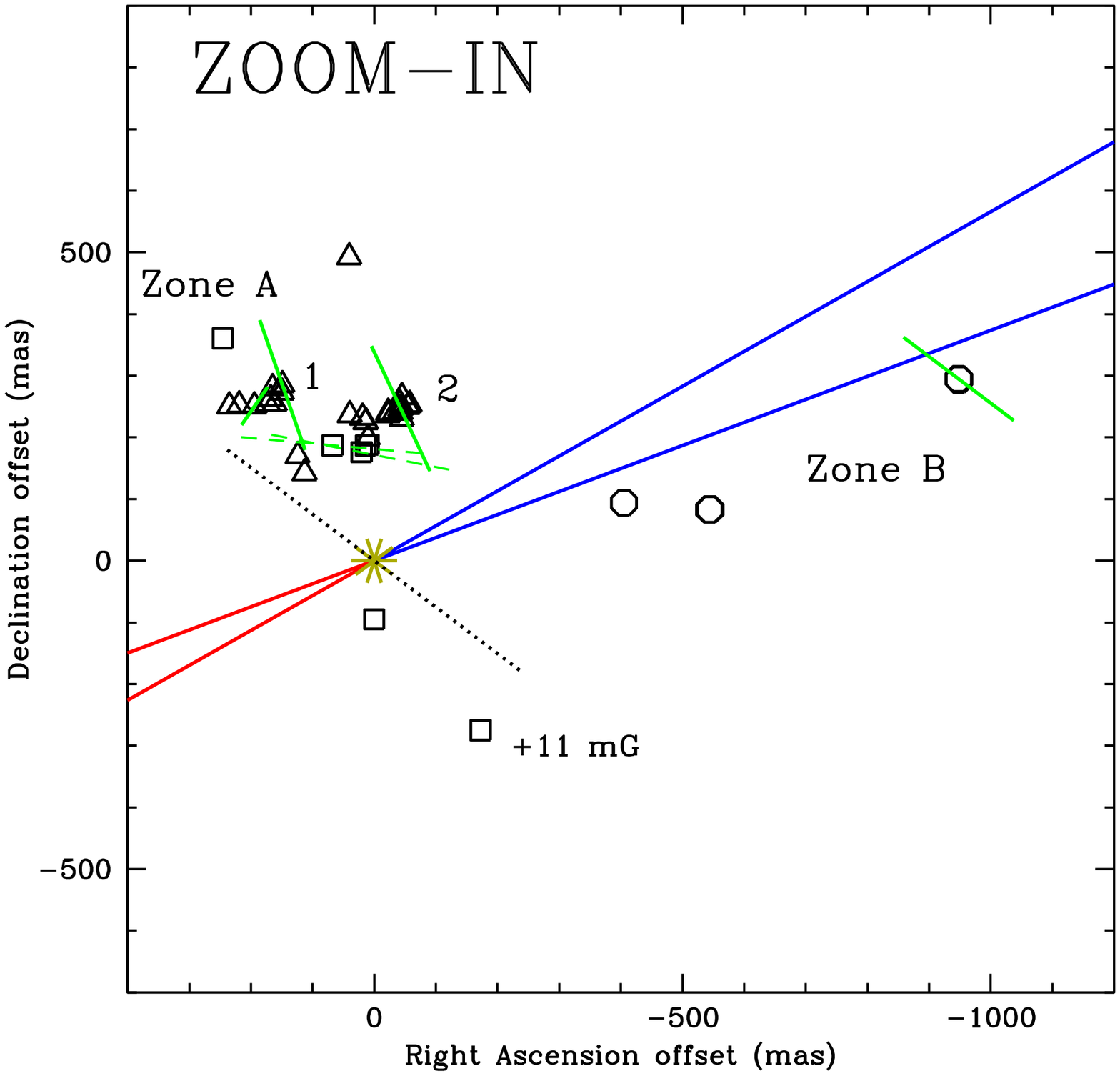}
\caption{Left panel: modified version of Fig.~3(b) of Shinnaga et al. (\cite{shi12}). The white box indicates the position of the right panel. The black bars 
represent the magnetic field direction determined from the polarized dust emission at 350~$\mu$m, whose continuum emission is in the background. 
Right panel: \meth ~(triangles), OH (squares) (Edris et al. \cite{edr05}), and \water ~(octagons) masers in IRAS20126+4104. The gold asterisk
 represents the B0.5 protostar 
($\alpha_{2000}=20^{\rm{h}}14^{\rm{m}}26^{\rm{s}}\!.0498$ and $\delta_{2000}=41^{\circ}13'32''\!\!.443$, MCR11), while the dotted line represents the
 Keplerian disk of $\sim$1000~au ($\rm{PA_{disk}}=53^{\circ}\pm7$\d, Cesaroni et al. \cite{ces05}). The red and blue lines indicate the red- and 
blue-shifted lobes of the jet, respectively, with a $\rm{PA_{\rm{jet}}}=115$\d ~and an opening angle of $9$\d ~(MCR11). The thick green segments represent the 
magnetic field direction determined from the polarized \meth ~and \water ~maser emissions. The green dashed segments represent the magnetic field direction
determined from the linearly polarized emission of OH masers (Edris et al. \cite{edr05}). The foreground Faraday rotation at 1.6-GHz is probably not
negligible and needs to be taken into account when interpreting the image (see Sect.~\ref{faraday}).
}
\label{MF}
\end{figure*}
\subsection{Faraday rotation}
\label{faraday}
The interstellar medium (ISM) between IRAS20126+4104 and the observer causes a rotation of the linear polarization vectors known as foreground 
Faraday rotation ($\Phi_{\rm{f}}$). Even if previous works (e.g., Surcis et al. \cite{sur11a}, \cite{sur12}, \cite{sur13}) have shown that this rotation 
is small at both 6.7-GHz and 22-GHz and do not affect the measurements of the magnetic field orientation, it is important to determine $\Phi_{\rm{f}}$ 
for IRAS20126+4104. The foreground Faraday rotation is given by
\begin{equation}
\Phi_{\rm{f}}[^{\circ}]=4.22\times10^{6}~\left(\frac{D}{[\rm{kpc}]}\right)~
\left(\frac{n_{e}}{[\rm{cm^{-3}}]}\right)~\left(\frac{B_{||}}{[\rm{mG}]}\right)~
\left(\frac{\nu}{[\rm{GHz}]}\right)^{-2},
\label{fari}
\end{equation}
where $D$ is the length of the path over which the Faraday rotation occurs, $n_{\rm{e}}$ and $B_{||}$ are the 
average electron density and the magnetic field along this path, respectively, and $\nu$ is the frequency. By assuming that the interstellar electron density, magnetic field, and distance are $n_{\rm{e}}\approx0.012\,\rm{cm^{-3}}$, $B_{||}\approx2\,\rm{\mu G}$ (Sun et al. \cite{sun08}),
and $D=1.64$~kpc, respectively,  $\Phi_{\rm{f}}$ is estimated to be 4\d$\!\!.0$ at 6.7-GHz and 0\d$\!\!.3$ at 22-GHz, but for 1.6-GHz 
OH masers $\Phi_{\rm{f}}\approx60$\d.\\
\indent Surcis et al. (\cite{sur12, sur13}) found that the linear polarization vectors of 6.7-GHz \meth ~masers are quite accurately aligned in all
the young stellar objects (YSOs) that they observed, indicating that the internal Faraday rotation ($\Phi_{\rm{i}}$) is negligible. In the case of
 22-GHz \water ~masers, $\Phi_{\rm{i}}$ is found to be negligible only if the \water ~masers are pumped by a C-shock (Kaufman \& Neufeld \cite{kau96}).
\subsection{Morphology of the magnetic field}
\label{magori}
The two maser species that are associated with two different structures of the YSO (i.e., the disk and the outflows, see Sect.~\ref{res}) probe the 
morphology of the magnetic field in two different zones of the protostar. The magnetic field close to the disk (Zone~A, at $\sim$400~au from 
the protostar), which is probed by the \meth ~masers, has an orientation on the plane of the sky of $\Phi_{\rm{B}}^{\rm{disk}}=20^{\circ}\pm16^{\circ}$, 
while close to the jet (Zone~B, at $\sim$1600~au from the protostar), which is probed by the 
\water ~masers, $\Phi_{\rm{B}}^{\rm{outflow}}=53^{\circ}\pm13^{\circ}$ (see Fig.~\ref{MF}). A comparison of the morphology of the magnetic field with the
 structure of the protostar reveals that the magnetic field is parallel to the disk ($\rm{PA_{disk}}=53^{\circ}\pm7^{\circ}$; Cesaroni et al. \cite{ces06})
 in Zone~B, and it rotates clockwise by 33\d ~in Zone~A, i.e., at $\sim400$~au from the central protostar. Here the magnetic field is perpendicular to the jet
 ($\rm{PA_{jet}}=115$\d; MCR11). Moreover, the angle between the magnetic field and the line of sight is 
$\langle\theta\rangle_{\rm{CH_3OH}}=82^{\circ +8^{\circ}}_{-41^{\circ}}$ in Zone~A and   
$\langle\theta\rangle_{\rm{H_2O}}=90^{\circ +9^{\circ}}_{-9^{\circ}}$ in Zone~B; i.e., the magnetic field is on the plane of the sky. Even if the 
magnetic field is not parallel
 to the jet, $\langle\theta\rangle_{\rm{CH_3OH}}$ is consistent with the inclination of the jet with respect to the line of sight, which is $\varphi=80$\d ~(MCR11). 
In addition, because $\Delta V_{\rm{Z}}$ is negative, the magnetic field in Zone~A is pointing 
towards the observer (e.g., Surcis et al. \cite{sur11b}). We note that Edris et al. (\cite{edr05}) identified  one Zeeman pair of OH masers, which 
indicates a magnetic field strength of about +11 mG in the direction pointing away from the observer at the opposite side of the disk from Zone~A 
(see Fig. \ref{MF}). Therefore, this could be evidence for the reversal of the magnetic field from above to below the disk. \\
\indent  Shinnaga et al. (\cite{shi12}) measured an \textit{S}-shaped morphology of the magnetic field on a large scale by observing the 
polarized dust emission at 350~$\rm{\mu m}$ (see Fig.~\ref{MF}; angular resolution $9''$, which at 1.64~kpc corresponds to $\sim15000$~au). They determined 
that the magnetic field changes its direction from N-S to E-W inside the infalling region ($r<0.1~\rm{pc}\approx20000~\rm{AU}$). 
The orientation of the magnetic field determined from the linearly polarized emission of \meth ~and \water ~masers is in good agreement with the large-scale magnetic field. The orientation of the magnetic field measured from the OH masers by Edris et al. (\cite{edr05}) suffers from a large uncertainty
due to the large foreground Faraday rotation. Because the OH masers arise in the same projected area of the \meth ~masers, for which $\Phi_{\rm{f}}$ is small,
the orientation of the magnetic field measured from both maser species could be expected to be the same. This implies that the magnetic field vectors of
OH masers should be rotated of approximately 60\d ~to be consistent with those of the \meth ~masers. This rotation is equal to the foreground Faraday rotation
estimated in Sect.~\ref{faraday}. Consequently, the magnetic field derived from the OH maser emission would also be consistent with the \textit{S}-shaped 
morphology measured by Shinnaga et al. (\cite{shi12}).\\
\indent The good agreement of the magnetic field from small to large scale suggests that the \meth ~masers of Group~1 are not on the disk but they are  
likely to be tracing material that is being accreted onto the disk along the magnetic field line as in Cepheus~A (Vlemmings et al. \cite{vle10}). 
Indeed, if the \meth ~masers of Group~1 were on the disk, we would have expected a resulting magnetic field that is much more random because of turbulent 
motions in the disk (Seifried et al. \cite{sei12b}).
The \meth ~masers of Group~2 are instead interpreted as tracing the material in the disk winds that is flowing out along the twisted magnetic field 
lines. In this case, the \meth ~masers should have a helical motion, like the SiO masers in Orion 
(Matthews et al. \cite{mat10}), which is consistent with the proper motion of Group~1 measured by MCR11.
\subsection{Role of the magnetic field}
\label{role}
\begin{figure}[t!]
\centering
\includegraphics[width = 12 cm]{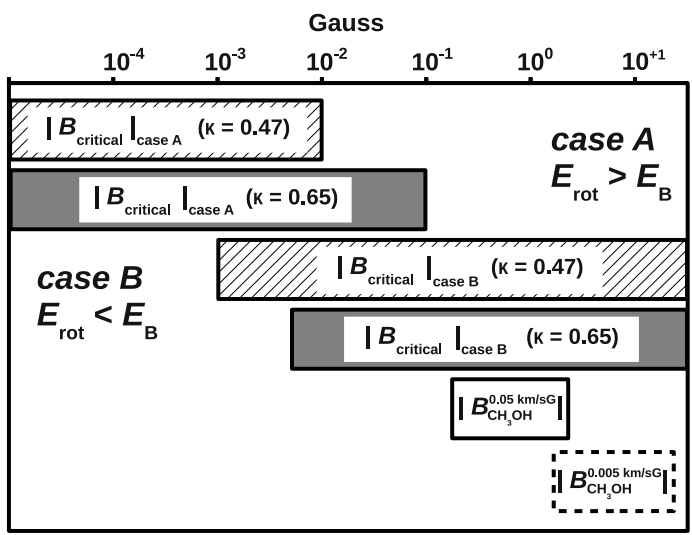}
\caption{A comparison between the magnetic field strength estimated from the Zeeman splitting of the \meth ~maser feature M05 
and the critical magnetic field strength if $E_{\rm{rot}}>E_{\rm{B}}$ and if 
$E_{\rm{rot}}<E_{\rm{B}}$. The hatched area and the dark gray area 
show the critical magnetic field strength in case $\kappa=0.47$ (Crutcher \cite{cru99}) and in case $\kappa=0.65$ (Crutcher et al. \cite{cru10}),
respectively. 
The full box indicates the range of $|B_{\rm{CH_3OH}}|$ measured by considering $\alpha_{\rm{Z}}^{\rm{CH_3OH}}=0.05$~\kmsg, while the dashed box indicates
the range of $|B_{\rm{CH_3OH}}|$ if $\alpha_{\rm{Z}}^{\rm{CH_3OH}}=0.005$~\kmsg. The ranges are estimated considering both the errors of \dvz ~and $\theta$.
}
\label{diagmeth}
\end{figure}

\indent To investigate the \textit{S}-shaped morphology Shinnaga et al. (\cite{shi12}) calculated the evolution of a magnetized cloud that has the same 
observed parameters of IRAS20126+4104. 
They considered a constant magnetic field strength of $1.5 \times 10^{-5}$~G parallel to the \textit{z} axis and with the rotation axis, which is 
rotated at an angle of 60\d ~with respect to the \textit{z} axis, on the \textit{y}-\textit{z} plane. In their simulations the initial cloud has the energy ratios
$E_{\rm{rot}}/E_{\rm{grav}}=0.02$ and $E_{\rm{B}}/E_{\rm{grav}}=0.55$, i.e. $E_{\rm{rot}}<E_{\rm{B}}$. Here $E_{\rm{rot}}$ is the 
rotational energy, $E_{\rm{grav}}$  the gravitational energy, and $E_{\rm{B}}$  the magnetic energy in the cloud. They find that
 the simulated magnetic field vectors agree with the 
observed morphology of the magnetic field if the cloud is observed from the \textit{x}-\textit{y} plane with a viewing angle of 30\d ~with respect to the \textit{y} axis.
More recently, Kataoka et al. (\cite{kat12}) have shown that in star-forming cores the polarization distribution  projected on the celestial plane
strongly depends on the viewing angle of the cloud. \\
\indent Kataoka et al. (\cite{kat12}) studied four different models in which they adopted a uniform magnetic field that has the same direction
but different strengths for each model. In Models~3 and 4, the rotation of the cloud is introduced and the rotation axis is inclined from the magnetic field
lines at an angle of 60\d. Model 4 has the strongest magnetic field among all the models. According to their simulations, the large-scale \textit{S}-shaped 
morphology, i.e. the magnetic field deviating from an hourglass configuration, in IRAS20126+4104 might be explained by Model~3, and it is
caused by (1) the misalignment of the magnetic field with the rotation axis and by (2) $E_{\rm{rot}}>E_{\rm{B}}$.  A slight misalignment of the magnetic field with 
the rotation axis was observed on a large scale by
 Shinnaga et al. (\cite{shi12}), who measured that the mean direction of the global magnetic field is $\rm{\Phi}_{\rm{B}^{\rm{global}}}=-3^{\circ}$, and the 
rotation axis of the cloud is $\rm{PA}_{\rm{rot}}=-40^{\circ}\pm20^{\circ}$. Condition (2) of Kataoka et al. (\cite{kat12}) instead contradicts 
the initial conditions of the simulations made by Shinnaga et al. (\cite{shi12}).\\
\indent So far, no observational determinations of the ratio between $E_{\rm{rot}}$ and $E_{\rm{B}}$ has been possible because no magnetic field strength has been 
measured in IRAS20126+4104. But now we can determine if $E_{\rm{rot}}>E_{\rm{B}}$ (hereafter case~A) or if $E_{\rm{rot}}<E_{\rm{B}}$ (hereafter case~B) by using our 
estimates of the magnetic field strength at \meth ~maser densities.\\
\indent  We assume that the cloud is a homogeneous solid sphere with magnetic flux freezing during its evolution. The rotational energy for a homogeneous solid 
sphere with radius $R$, mass $M$, and angular velocity $\Omega$ is
\begin{equation}
E_{\rm{rot}}=\frac{1}{5} MR^2\Omega^2,
\label{erot}
\end{equation}
while the magnetic energy for the same sphere is
\begin{equation}
E_{\rm{B}}=\frac{1}{6}|B|^2 R^3,
\label{eb}
\end{equation}
where $|B|$ is the magnetic field strength into which the sphere is immersed. The critical value of magnetic field at which $E_{\rm{rot}}=E_{\rm{B}}$ is
\begin{equation}
|B_{\rm{critical}}|=\sqrt{\frac{6}{5}} \cdot \frac{M^{1/2} ~\Omega}{R^{1/2}}.
\label{B}
\end{equation}
Considering that the estimates for the cloud properties of IRAS20126+4104 are $R=0.54~\rm{pc}$, 
$M=402$~\solmass ~(Hofner et al. \cite{hof07}), and $\Omega=2~\rm{km~s^{-1}~pc^{-1}}$
(Shinnaga et al. \cite{shi08}), we find that the critical value of the magnetic field of the cloud should be
\begin{equation}
|B_{\rm{critical}}|=5\times 10^{-5}~\rm{G}.
\label{B_theor}
\end{equation}
This $|B_{\rm{critical}}|$ value is determined not at the \meth ~maser densities,  so it cannot be directly compared with the magnetic field
strength measured by us. But because we have assumed the presence of magnetic flux freezing in the cloud, the relation $|B|\propto n_{\rm{H_2}}^{\kappa}$,
where $\kappa=0.47$ as empirically determined by Crutcher (\cite{cru99}), can be used to estimate $|B_{\rm{critical}}|$ at the 
\meth ~maser densities. We assume $|B|\propto n_{\rm{H_2}}^{0.47}$ because it is proven to be valid up to densities of $10^{11}~\rm{cm^{-3}}$ (Vlemmings \cite{vle08}).
Cragg et al. (\cite{cra05}) determine that the number density of 6.7-GHz \meth ~maser ($n_{\rm{H_2}}^{\rm{CH_3OH}}$) varies from
 $10^{7}~\rm{cm^{-3}}$ to $10^{9}~\rm{cm^{-3}}$, above which the \meth ~masers are quenched. Therefore,  we have to estimate a range of $|B_{\rm{critical}}|$ by  
 considering the whole range of $n_{\rm{H_2}}^{\rm{CH_3OH}}$. The critical value of the magnetic field at the densities of the 
6.7-GHz \meth ~maser is thus between $|B_{\rm{critical}}^{10^7~\rm{cm^{-3}}}|=0.001~\rm{G}$ and $|B_{\rm{critical}}^{10^9~\rm{cm^{-3}}}|=0.01~\rm{G}$. Consequently, in 
Case~A, $|B_{\rm{critical}}^{10^9~\rm{cm^{-3}}}|_{\rm{case A}}<0.01~\rm{G}$ ($\kappa=0.47$), and in Case B, 
$|B_{\rm{critical}}^{10^7~\rm{cm^{-3}}}|_{\rm{case B}}>0.001~\rm{G}$ ($\kappa=0.47$). \\
\indent It is important to mention that Crutcher et al. (\cite{cru10}) claim a different value of $\kappa$, i.e. $\kappa=0.65$. They find that at densities less 
than $10^2~\rm{cm^{-3}}$,
magnetic fields are density independent; i.e., they are constant, while for higher densities they vary as $|B|\propto n_{\rm{H_2}}^{0.65}$. Even though this relation has so far been verified for densities up to $10^{7}~\rm{cm^{-3}}$, for the sake of completeness we also estimate $|B_{\rm{critical}}|$ at \meth ~maser densities 
by using $|B|\propto n_{\rm{H_2}}^{0.65}$. Repeating the calculation for $\kappa=0.65$, we found
$|B_{\rm{critical}}^{10^9~\rm{cm^{-3}}}|_{\rm{case A}}<0.1~\rm{G}$ ($\kappa=0.65$) and  
$|B_{\rm{critical}}^{10^7~\rm{cm^{-3}}}|_{\rm{case B}}>0.004~\rm{G}$ ($\kappa=0.65$).\\
\indent In Fig.~\ref{diagmeth} we show a simple diagram that can help visualize the different $|B_{\rm{critical}}|$ ranges and the measured $|B_{\rm{CH_3OH}}|$, which are 
estimated by using both $\alpha_{\rm{Z}}=0.05$~\kmsg ~and 
$\alpha_{\rm{Z}}=0.005$~\kmsg. To determine the ranges of $|B_{\rm{CH_3OH}}|$, we also considered the errors of \dvz ~and $\theta$. 
We can see from Fig.~\ref{diagmeth} that the magnetic field measured from the Zeeman splitting of the \meth ~maser M05, independently of the value 
of $\alpha_{\rm{Z}}$ and $n_{\rm{H_2}}^{\rm{CH_3OH}}$, indicates that $E_{\rm{rot}}<E_{\rm{B}}$ (both for $\kappa=0.47$ and for $\kappa=0.65$).\\
\indent Using similar calculations for 1.6-GHz OH maser ($10^5~\rm{cm^{-3}}<n_{\rm{H_2}}<10^8~\rm{cm^{-3}}$; Crutcher \cite{cru12}), we find that the magnetic field strength measured 
by Edris et al. (\cite{edr05}), i.e. 11~mG, satisfies Case B, i.e. $E_{\rm{rot}}<E_{\rm{B}}$, only if $\kappa=0.47$ 
($|B_{\rm{critical}}^{10^8~\rm{cm^{-3}}}|_{\rm{case B}}>10^{-4}$~G) and Case A, i.e. $E_{\rm{rot}}>E_{\rm{B}}$, only if $\kappa=0.65$ 
($|B_{\rm{critical}}^{10^5~\rm{cm^{-3}}}|_{\rm{case A}}<0.02$~G).\\
\indent Therefore, in our estimates the magnetic field dominates the rotation of the cloud. Moreover, we can speculatively state that the initial conditions of Shinnaga 
et al. (\cite{shi12}) are correct and that the $S$-shaped morphology of the magnetic field cannot be described by Model~3 of Kataoka et al. (\cite{kat12}). 
However, in Model~4 of Kataoka et al. (\cite{kat12}), the magnetic field is stronger, and we have the initial condition $E_{\rm{rot}}<E_{\rm{B}}$. In this case they find
that the deviation of the magnetic field lines from the hourglass configuration could only be observed very close to the protostar, i.e., 
 where the magnetic field is probed by the 6.7-GHz \meth ~masers.
Of course, further observations, for instance of dust tracers in full polarization mode at mas resolution, could in future help clarify the role of the magnetic 
field in IRAS20126+4104.
\section{Conclusions}
The YSO IRAS20126+4104 has been observed in full polarization spectral mode at 6.7-GHz with the EVN and at 22-GHz with the VLBA to detect
linear and circular polarization emission from \meth ~and \water ~masers, respectively. We detected 26 \meth ~masers and 5 \water ~masers at mas resolution.
Linearly polarized emission was detected towards three \meth ~masers and one \water ~maser that probed the magnetic field both close to the Keplerian disk and
to the large-scale outflow. The orientation of the magnetic field derived from the masers agrees with the $S$-shaped morphology that was measured 
by Shinnaga et al. (\cite{shi12}) on a larger scale by using dust-polarized emission at 350~$\mu$m.\\
\indent Moreover, we were able to measure a Zeeman splitting of -9.2~\ms ~from the brightest 6.7-GHz \meth ~maser. From this measurement, we determined that the 
magnetic field energy dominates the rotation energy of the region; i.e., $E_{\rm{rot}}<E_{\rm{B}}$.\\

\begin{acknowledgements}
We wish to thank an anonymous referee for making useful suggestions that have improved the paper.
 The EVN is a joint facility of European, Chinese, South 
African, and other radio astronomy institutes funded by their national research councils.
 \end{acknowledgements}
\bibliographystyle{aa} 

\end{document}